\documentclass[usenatbib]{mnras}

\usepackage{newtxtext,newtxmath}
\usepackage[T1]{fontenc}
\usepackage{ae,aecompl}

\usepackage{graphicx}	% Including figure files
\usepackage{amsmath}	% Advanced maths commands
\usepackage{amssymb}	% Extra maths symbols

\title[Light bosonic dark matter with SDSS Lyman-$\alpha$ forest]{Constraining the mass of light bosonic dark matter using SDSS Lyman-$\alpha$ forest.}

\author[E. Armengaud et al.]{
Eric Armengaud,$^{1}$\thanks{E-mail : eric.armengaud@cea.fr}
Nathalie Palanque-Delabrouille,$^{1}$
Christophe Yèche,$^{1,2}$
\newauthor David J. E. Marsh,$^{3}$
Julien Baur~$^{1}$
\\
$^{1}$IRFU, CEA, Université Paris-Saclay, F-91191 Gif-sur-Yvette, France\\
$^{2}$Lawrence Berkeley National Laboratory, Berkeley, CA 94720, USA\\
$^{3}$Department of Physics, King's College London, Strand, London, WC2R 2LS, UK}
\date{Accepted XXX. Received YYY; in original form ZZZ}
\pubyear{2017}

\begin{document}
\label{firstpage}
\pagerange{\pageref{firstpage}--\pageref{lastpage}}
\maketitle

\begin{abstract}
If a significant fraction of the dark matter in the Universe is made of an ultra-light scalar field, named fuzzy dark matter (FDM) with a mass $m_a$ of the order of $10^{-22}-10^{-21}$~eV, then its de Broglie wavelength is large enough to impact the physics of large scale structure formation. In particular, the associated cutoff in the linear matter power spectrum modifies the structure of the intergalactic medium (IGM) at the scales probed by the Lyman-$\alpha$ forest of distant quasars. We study this effect by making use of dedicated cosmological simulations which take into account the hydrodynamics of the IGM. We explore heuristically the amplitude of quantum pressure for the FDM masses considered here and conclude that quantum effects should not modify significantly the non-linear evolution of matter density at the scales relevant to the measured Lyman-$\alpha$ flux power, and for $m_a \ga 10^{-22}$~eV. We derive a scaling law between $m_a$ and the mass of the well-studied thermal warm dark matter (WDM) model that is best adapted to the Lyman-$\alpha$ forest data, and differs significantly from the one inferred by a simple linear extrapolation. By comparing FDM simulations with the Lyman-$\alpha$ flux power spectra determined from the BOSS survey, and marginalizing over relevant nuisance parameters, we exclude FDM masses in the range $10^{-22} \la m_a < 2.3\times 10^{-21}$~eV at 95~\% CL. Adding higher-resolution Lyman-$\alpha$ spectra extends the exclusion range up to $2.9\times 10^{-21}$~eV. This provides a significant constraint on FDM models tailored to solve the "small-scale problems" of $\Lambda$CDM.
\end{abstract}

\begin{keywords}
large-scale structure of Universe -- dark matter
\end{keywords}

\section{Introduction}

While non-baryonic dark matter is a cornerstone in our current understanding of the observable Universe, its nature remains completely unknown after decades of theoretical and experimental investigations. Cold dark matter (CDM) is necessary to understand at the same time the global dynamics of the Universe and the gravitational formation of large scale structures from an initial, almost scale-invariant, power spectrum. The current-time dark matter density derived from CMB observations, $\Omega_c h^2 = 0.118 \pm 0.001$~\citep{PlanckCollaboration2015} is six times larger than the baryon density $\Omega_b h^2 = 0.0222 \pm 0.0002$, which can also be independently derived from the measurements of light element abundances. Dark matter also seems to be necessary to understand the dynamics of gravitationally bound systems such as galaxies~\citep{Persic1996} and clusters of galaxies~\citep{Clowe2006}.

To understand the nature of dark matter, two complementary strategies are possible. A first option is to test specific models inspired by particle physics, using more or less direct detection strategies. Two outstanding candidates predicted in extensions of the Standard Model of particle physics are the supersymmetric neutralino, a thermal relic whose mass is in the GeV$-$TeV range, and the QCD axion, a relic pseudo-scalar field with a mass around $1-100\,\umu\rm{eV}$. They are tested with dedicated experiments or observations, without a positive outcome as of now~\citep{Olive2016a} although future experiments will explore much more of their parameter space, in particular in the case of the currently poorly tested QCD axion.
A second option consists in measuring or constraining generic properties of dark matter from astrophysical or cosmological observations. For example, assuming that the dark matter is made of fermions, then estimations of their phase-space density in dwarf galaxies imply the Tremaine-Gunn bound~\citep{Tremaine1979a} on their mass, $m \gtrsim 400$~eV. Such a bound does not apply for bosonic particles. 

The small-scale clustering features of visible matter may also be used in different ways to constrain dark matter properties. In fact, at the galactic or sub-galactic scale it has been argued that CDM predictions do not match observations, concerning the number density of low-mass and dwarf galaxies~\citep{Klypin1999, Moore1999, Boylan-Kolchin2012}, and the absence of an observed central dark matter cusp in these objects~\citep{Oh2011}. While models which take into account the full richness of baryon physics at these scales may surpass those issues, it is still possible that these features result from specific DM properties, not included in the most naive CDM model.

A popular scenario to explain the apparent lack of structures at small scales is that the DM velocity dispersion could be large enough at the time of equality, when structures start to form, so that the related free-streaming of these warm DM (WDM) particles would suppress small-scale structures. For thermal WDM of mass $m_X$, using cosmological parameters from~\citet{PlanckCollaboration2015}, the associated cutoff in the linear density power spectrum $P(k)$ is at the scale~\citep{Viel2005} :

\begin{equation}
k_c = 6.46\,\left(\frac{m_X}{\rm keV}\right)^{1.11} \, {\rm Mpc}^{-1} \;\;{\rm where}\;\; \sqrt{\frac{P_{\rm WFM}(k_c)}{P_{\rm CDM}(k_c)}}=0.5
\label{eqn:kc_wdm}
\end{equation}

\noindent In this equation, $k_c$ does not depend on $h$. Recent observations of a 3.5~keV X-ray line in several DM-rich objects~\citep{Bulbul2014,Cappelluti2017a} were also interpreted as a hint for a 7~keV sterile neutrino which would constitue WDM. An issue with WDM is that it seems difficult to solve the halo abundance and the cusp-core problems at the same time, if both are to be taken seriously. This is the "Catch 22" problem~\citep{Maccio2012} : for WDM with a mass of a few keV, the halo core size, which originates from the above-mentioned Tremaine-Gunn phase-space density argument, is too small to solve the apparent cusp-core issue.

Other DM properties were suggested and studied to alleviate the small-scale CDM issues. One possibility is that DM is self-interacting with large cross-sections of the order of $0.1 - 1$~cm$^2$/g, thus producing constant-density cores inside halos~\citep{Spergel2000, Rocha2012}. The free-streaming effect characteristic of WDM could also be produced by decaying DM with a small mass splitting~\citep{Wang2013}.

In this article, we now focus on the alternative scenario according to which DM consists in extremely light bosons, with a mass scale $m_a\sim 10^{-22}$~eV, so that the quantum wave properties of these particles are large enough to smooth the associated density fluctuations on the relevant small scales. The basic physics involved at large scales in this Fuzzy Dark Matter (FDM) scenario is presented in eg.~\citet{Hu2000a, Marsh2016a, Hui2016}. The effective "quantum pressure" associated with FDM generates an effective Jeans wavenumber, which is the following for a fluid of density $\rho$ :

\begin{eqnarray}
k_J(\rho) & =  & 2\left(\upi G \rho\right)^{1/4}\,\left(\frac{m_a}{\hbar}\right)^{1/2} \\
 & = & \frac{0.11}{h^{1/2}}\,\left(\frac{\rho}{\rho_c}\right)^{1/4}\,\left(\frac{m_a}{10^{-22}\,{\rm eV}}\right)^{1/2}\;{\rm kpc}^{-1}
\end{eqnarray}

Structure formation is suppressed for comoving scales smaller than the corresponding Jeans scale at equality time, which is roughly $\sim 10$~Mpc$^{-1}$ for $m_a=10^{-22}$~eV. This results in a linear power spectrum with a small-scale cutoff produced at high redshift. The related phenomenology is therefore close to the WDM one. On the other hand, in dense environments such as the center of DM halos, where the effective Jeans scale is small due to the $\rho^{1/4}$ scaling, simulations which take explicitly into account these wave properties~\citep{Woo2009, Schive2014, Veltmaat2016, Zhang2016a} demonstrate a rich phenomenology including the development of solitonic cores inside Navarro-Frenk-White halos. In particular, these solitonic cores can fit the observed kinematic properties of some dwarf galaxies, with FDM masses ranging from $\sim 0.4\times 10^{-22}$~eV up to $\sim 4 - 6 \times 10^{-22}$~eV depending on the considered objects and analysis procedure~\citep{Schive2014, Marsh2015, Calabrese2016a, Chen2016, Gonzales-Morales2016}.

There is to our knowledge no strong argument to explain why the DM mass would be such that their de Broglie wavelength is of the order of galactic scales. On the other hand, light bosonic pseudo-scalars are a generic prediction in string models, for example as the zero-mode of the Kaluza-Klein tower resulting from the compactification of antisymmetric tensor fields, with the mass scale arising due to the modulus fixing the GUT scale coupling~\citep{Svrcek2006}. It is plausible that DM is made of several such species, one of which would have the right mass to produce the FDM phenomenology. However for simplicity, we consider in this article only the case of a single (fuzzy) DM species.

Existing bounds on the mass of FDM particles arise from both linear and non-linear cosmological probes. The most robust contraints are based on linear probes of structure formation, mostly the CMB, which show that FDM is not a dominant component of dark matter for $m_a \lesssim 10^{-25}$~eV~\citep{Hlozek2015}. Models with higher values for $m_a$ cannot be distinguished from CDM given that the scales for which structure formation is damped with respect to CDM are too small to be observed with the primary CMB anisotropies. Pushing into the mildly non-linear regime, CMB lensing at high $\ell$ could explore FDM masses up to $\sim 10^{-23}$~eV~\citep{Hlozek2016}. For even higher-mass FDM, the halo mass function (HMF) was semi-analytically computed from the linear power spectrum, using formalisms similar to the case of pure CDM~\citep{Marsh2014a, Marsh2016b, Du2017} : for the benchmark mass $m_a=10^{-22}$~eV, the HMF starts to be suppressed for $M \la 10^{10}\,M_{\sun}$, while it is fully cut off at $M\sim 10^{7} - 10^{8}\,M_{\sun}$ depending on the calculation. The galaxy UV-luminosity function at high redshift $z\sim 4-10$ is therefore expected to be strongly attenuated for $M_{\rm UV} \ga -16$, and accordingly reionization is also expected to be delayed~\citep{Bozek2014, Sarkar2016}. Recent estimations of the high-redshift UV-luminosity function~\citep{Bouwens2015, Livermore2016}, as derived from deep observations such as the Hubble Frontier Fields, are still subject to large uncertainties~\citep{Bouwens2016} but they do not clearly point towards a deviation from the CDM Schechter law. FDM bounds could therefore be derived, ranging from $m_a > 1.2\times 10^{-22}$~eV to $5\times 10^{-22}$~eV or even $1.2\times 10^{-21}$~eV, depending on the data used and analysis strategy~\citep{Schive2015, Corasaniti2016, Menci2017}.

The absorption of the Lyman-$\alpha$ emission from distant quasars by neutral \ion{H}{I} in the intergalactic medium (IGM) constitutes a powerful, high-redshift probe for spatial fluctuations of the matter density at comoving scales going down to $\sim 1 - 0.1$~Mpc depending on the instrumental wavelength resolutions. As in the case of WDM, it is therefore expected that Lyman-$\alpha$ forest observations can constrain FDM models, with very different sources of uncertainties with respect to the aforementioned measurements. In fact, \citet{Amendola2006a} disfavors pure FDM models with $m_a < 10^{-22}$~eV, by making use essentially of high-resolution Lyman-$\alpha$ forest spectra. Given the similar phenomenologies of FDM and WDM, it is also possible to apply existing WDM bounds on $m_X$ (eg.~\citet{Viel2013, Baur2015}) to the FDM scenario, by using the correspondance between $m_a$ and $m_X$ which matches the cutoff position $k_c$ in the linear matter power spectrum~\citep{Marsh2014a}. This correspondance can roughly be estimated as highlighted in~\citet{Marsh2016a} : the power spectrum suppression for WDM takes place for modes inside the horizon when $T\sim m_X$, while for FDM the suppressed modes are inside the horizon when the scalar field starts oscillating, i.e. for $H(T)\sim m_a$. Since $T\sim \sqrt{H M_{\rm Pl}}$ (radiation era), the order-of-magnitude mass matching between both models is thus

\begin{equation}
\frac{m_X}{\rm keV} \sim \frac{\sqrt{m_a\,M_{\rm Pl}}}{\rm keV} \sim 0.5 \left( \frac{m_a}{10^{-22}\,{\rm eV}} \right)^{0.5}
\label{eqn:mass-scaling}
\end{equation}

This relation is confirmed by more sophisticated calculations. The bound $m_X > 4.09$~keV~\citep{Baur2015} would therefore naively translate into $m_a \ga 7\times 10^{-21}$~eV. However the significant differences of power spectrum shapes between FDM and WDM at the cutoff scale, as computed in~\citet{Marsh2014a, Schive2015}, make this interpretation uncertain and, as we will see, too optimistic.

Given the potential of the Lyman-$\alpha$ forest to constrain FDM, it is desirable to dedicate more in-depth studies on this scenario. In this article we examine the impact of the exact FDM linear power spectrum on the Lyman-$\alpha$ forest. We present the results of hydrodynamical simulations which make use explicitly of FDM transfer functions. We highlight the impact of the difference between WDM and FDM linear transfer functions on the predicted Lyman-$\alpha$ flux spectra. While we do not take into account the full wave behavior of the FDM fluid in the numerical simulations, we present quantitative arguments which suggest that this approximation should be valid at the scales probed by our simulations for $m_a \ga 10^{-22}$~eV. We then combine our models with existing high-statistics Lyman-$\alpha$ power spectra obtained from the BOSS survey, as published in~\citet{Palanque-Delabrouille2013a}, to derive constraints on $m_a$, by applying a procedure similar to that used in e.g.~\citet{Baur2015} to take into account various sources of uncertainties. Finally, we also add higher-resolution Lyman-$\alpha$ spectra from~\citet{Lopez2016} and~\citet{Viel2013} to the analysis in order to improve the sensitivity to higher values of $m_a$.

\section{Calculation of the Lyman-$\alpha$ power spectrum}

\subsection{Linear matter spectrum}

Throughout this paper we consider for simplicity a pure-FDM model, in which all of the dark matter is made of a (pseudo) scalar field $\Psi$, of mass $m_a$. We also do not consider possible self-interactions. In the scenario when such interactions are given by the expansion of a periodic axion-like potential $V\sim f_a^2\,m_a^2(1-\cos \Psi/f_a)$, their effects are negligible at the cosmological scales of interest for values of $f_a$ that do not require fine-tuning to give the relic abundance. The homogeneous component $\Psi_0$ of the field obeys then the Klein-Gordon equation in the Friedman-Robertson-Walker metric:

\begin{equation}
\ddot{\Psi}_0 + 3 H(z)\,\dot{\Psi}_0 + m_a^2\,\Psi_0=0
\end{equation}

For the masses of interest here, $m_a \gg H(z)$ during all the epoch of structure formation so that the field oscillates and behaves on large scales as CDM. On small scales, the effect of the pressure associated with $\Psi$ gives to the FDM fluid an effective sound velocity (eg.~\citet{Hwang2009})

\begin{equation}
c_{\rm s,eff}^2 = \frac{k^2/4\,m_a^2}{1+k^2/4\,m_a^2}
\end{equation}

The related Jeans scale damps structure formation in the linear regime at small scales. Therefore the transfer function defined from the linear matter power spectrum $P_{\rm FDM}(k) = T^2(k) \times P_{\rm CDM}(k)$ presents a strong attenuation at the comoving Jeans scale during equality. We calculate $P_{\rm FDM}(k)$ using the \texttt{AxionCAMB}~\citep{Hlozek2016} modification of the \texttt{CAMB} Boltzmann solver~\citep{Lewis2000}, with all cosmological parameters from~\citet{PlanckCollaboration2015}.

\begin{figure}
\includegraphics[width=\columnwidth]{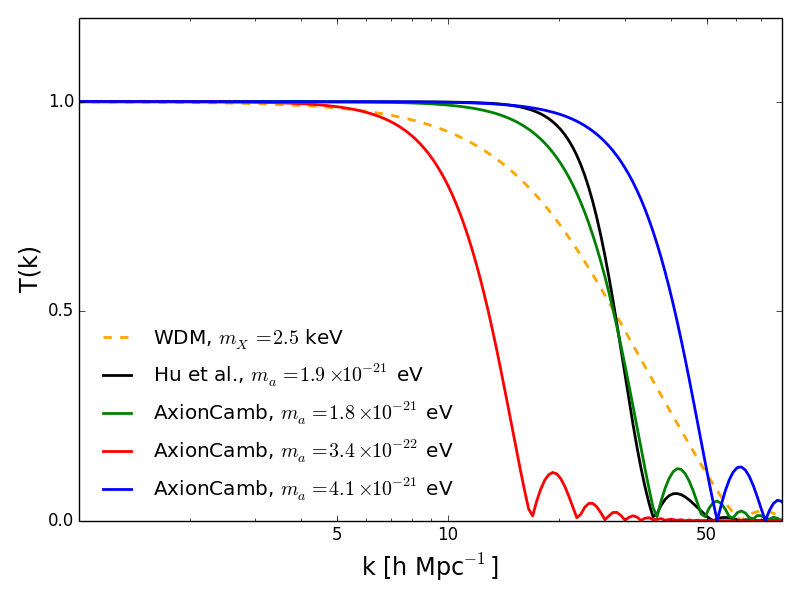}
\caption{The linear transfer function $T(k)$ with respect to pure cold dark matter, for several FDM models, as well as for a 2.5~keV WDM model computed with the \texttt{CLASS} Boltzmann solver~\citep{Lesgourgues2011}. The axion masses $m_a=1.9\times 10^{-21}$~eV (Hu et al. parametrization) and $1.8\times 10^{-21}$~eV (\texttt{AxionCAMB}) were chosen to match the WDM model when $T(k)=0.5$.}
\label{fig:tklin}
\end{figure}

Fig.~\ref{fig:tklin} illustrates the corresponding cutoff calculated for three axion masses in the range of interest for this study, namely $3\times 10^{-22} \la m_a \la 5\times 10^{-21}$~eV. We also show the FDM transfer function as parametrized by~\citet{Hu2000a}. For identical values of $m_a$, we observe a slight, but significant shape difference between this parametrization and \texttt{AxionCAMB} : the power spectrum cutoff  from the full \texttt{AxionCAMB} calculation is smoother than the one from~\citet{Hu2000a}. This trend is confirmed by~\citet{Urena-Lopez2016}, a \texttt{CLASS}-based~\citep{Lesgourgues2011} calculation of the FDM transfer function. Based on \texttt{AxionCAMB} calculations for several values of $m_a$ in the range $10^{-23} - 10^{-20}$~eV, we find the following scaling relation for the cutoff mode $k_c$ defined by $T(k_c)=0.5$ :

\begin{equation}
k_c = 4.97\,\left( \frac{m_a}{10^{-22}\,{\rm eV}} \right)^{0.46}\,{\rm Mpc}^{-1}
\label{eqn:fdm-cutoff}
\end{equation}

This relation is in good agreement with those found in~\citet{Hu2000a, Marsh2014a}. From Eqn.~\ref{eqn:kc_wdm}, the mass-matching relation between WDM and FDM becomes 

\begin{equation}
m_X = 0.79\,\left(\frac{m_a}{10^{-22}\,{\rm eV}}\right)^{0.42}\;{\rm keV}, 
\label{eqn:mass-scaling-2}
\end{equation}

\noindent again in reasonable agreement with previous estimates, including the order-of-magnitude value given in Eqn.~\ref{eqn:mass-scaling}.

The WDM transfer function, also represented on Fig.~\ref{fig:tklin} for $m_X=2.5$~keV, is much smoother than the FDM one. Therefore it is expected that SDSS-based constraints on FDM will differ significantly from those obtained by the mapping of Eqn.~\ref{eqn:mass-scaling-2}, which justifies the need for dedicated simulations. Given that the Lyman-$\alpha$ forest essentially probes the one-dimensional matter power spectrum,

\begin{equation}
P_{\rm 1D}(k) = \frac{1}{2\,\upi} \int_{k}^{\infty} d\tilde{k} \,\tilde{k}\,P_{\rm 3D}(\tilde{k}),
\end{equation}

\noindent the cutoff in the 3-D spectrum modeled by a given $T(k)$ will generically produce a cutoff in the 1-D spectrum with a different, smoother shape, and at much smaller values of $k$ (which explains why SDSS spectra, with $k\la 2\,h\,{\rm Mpc}^{-1}$, are sensitive to models with $k_c$ of several tens $h\,{\rm Mpc}^{-1}$). Non-linear and hydrodynamical effects make the picture more complicated, so that it is difficult to predict without full calculations how Lyman-$\alpha$ flux power spectra in the SDSS range are modified by changes in the shape of $T(k)$. It is nevertheless expected that SDSS Lyman-$\alpha$ data probe essentially the low-$k$ part of $T(k)$, and therefore constraints on $m_a$ should be less severe than those anticipated by making use of WDM bounds with relations such as Eqn.~\ref{eqn:mass-scaling-2}. Adding higher-resolution Lyman-$\alpha$ data will naturally extend the range of explored $k$, therefore improving the sensitivity to higher values of $m_a$. 

\subsection{Non linear matter spectrum}

In the non-linear, newtonian and non-relativistic regime i.e. at low redshifts and reasonably small scale, the evolution of the FDM fluid can be rigorously followed using two equivalent methods : either by explicitly solving the Poisson-Schr\"{o}dinger equations as in~\citet{Schive2014}, or by solving the corresponding Madelung fluid equations. Given the orders of magnitude for the DM density, velocity and de Broglie wavelength, the particle occupancy numbers are very high so that these equations can be interpreted in a classical way. For a FDM fluid whose wave function writes $\Psi = \sqrt{\rho} \, e^{i\theta}$, with $m\vec{v} = \hbar \nabla\theta$, the dynamical equation in a background geometry characterized by the scale factor $a$ and expansion rate $H$, with newtonian potential $\phi$, is (see e.g.~\citet{Chavanis2011}):

\begin{equation}
\upartial_{t} \vec{v} + H\,\vec{v} + \frac{1}{a} (\vec{v}\cdot\nabla)\vec{v} = -\frac{1}{a} \nabla \left[ \phi - \frac{\hbar^2}{2m_a^2\,a^2} \left( \frac{\nabla^2 \sqrt{\rho}}{\sqrt{\rho}}\right) \right]
\label{eqn:euler}
\end{equation}

This is equivalent to the Euler equation as implemented in standard N-body simulations, with an additional "quantum pressure" term $Q \equiv - (\hbar^2/2m_a) \frac{\nabla^2 \sqrt{\rho}}{\sqrt{\rho}}$. Note that, as~\citet{Veltmaat2016} show, solving this fluid equation gives identical results to the Poisson-Schr\"{o}dinger system away from interference fringes. Implementing this term into N-body simulations is difficult~\citep{Marsh2015a, Mocz2015} mostly because it is a second order derivative of the density field, which means that most of its power, in terms of Fourier transform, is on very small scales. In~\citet{Veltmaat2016}, the effect of quantum pressure in the non-linear regime was recently studied at cosmological scales using a particle-mesh scheme. For $m_a=2.5\times 10^{-22}$~eV, significant modifications of the matter power spectrum are visible for $k \ga 100$~Mpc$^{-1}$. This is consistent with arguments from~\citet{Schive2015} : simulations with $m_a \sim 10^{-22}$~eV have shown that the $Q$ term has the most severe impact on the internal structures of DM halos, and much less on larger scales. However, the SDSS Lyman-$\alpha$ spectra do not probe comoving scales smaller than a few hundreds of kpc. As another argument, we computed with \texttt{AxionCAMB} the linear growth rate ratio $\xi$ defined in~\citet{Schive2015} : for $m_a=3.4\times 10^{-22}$~eV, we find $\xi=1$ within 1~\% for $k\la 20\,h\,{\rm Mpc}^{-1}$. We therefore expect the $Q$ term to have a negligible influence on the DM dynamics with respect to gravity, at the scale relevant for the observable Lyman-$\alpha$ forest.

As a consequence, for this study it is most probably appropriate to use standard N-body simulations. In Section 3.1, we will show \textit{a posteriori} distributions for the $Q$ term which support this choice.

\subsection{Hydrodynamical simulations}

The methodology and numerical tools used here are very similar to and were extensively described in~\citet{Borde2014,Rossi2014a,Palanque-Delabrouille2015a,Baur2015}. We use the \texttt{Gadget-3} N-body simulation, an update of the \texttt{Gadget-2} program~\citep{Springel:2000yr,Springel2005}. In this exploratory work, which aims to assess specifically the impact of the shape of the linear power spectrum on the resulting Lyman-$\alpha$ flux, we use a single set of $\Lambda$CDM cosmological parameters from~\citet{PlanckCollaboration2013}. They are in accordance to the central values used in~\citet{Baur2015} : $h=0.675$, $\Omega_M=0.31$, $n_s=0.96$ and $\sigma_8=0.83$. Note that we checked that, as for WDM, the cutoff in the linear matter power spectrum induced by the FDM models considered here does not change significantly the value of $\sigma_8$, for a given primordial scalar perturbation amplitude.

The initial conditions are set at $z=30$ with the \texttt{2LPTIC} software, starting from the linear matter power spectrum as computed by \texttt{AxionCAMB} for this redshift. The (fuzzy) dark matter fluid is then treated as a collection of fixed-mass point particles. The baryon fluid is evolved using the Smoothed Particle Hydrodynamics technique, with stars created in cold and dense baryon environments. Of importance for the Lyman-$\alpha$ forest, we model the IGM heating by the UV background light, using internal \texttt{Gadget} heating rate parameters which result in the redshift-dependent IGM temperature-density relation 

\begin{equation}
T_{\rm IGM} = T_0(z) (1+\delta \rho/\rho)^{\gamma(z)-1}
\label{eqn:t_igm}
\end{equation}

\noindent As for the case of cosmological parameters, we adopt here fixed benchmark parameters such that $T_0(z=3)=14000$~K and $\gamma(z=3)=1.3$, which are in agreement with the measurements from~\citet{Becker2010}.

From \texttt{Gadget} snapshot files in the redshift range $z=4.6 - 2.2$ adapted to SDSS Lyman-$\alpha$ data, we infer the line-of-sight-averaged one-dimensional Lyman-$\alpha$ flux power spectrum. This observable is defined from the fluctuations of quasar's transmitted flux fraction, $\delta_\varphi(\lambda) = \varphi(\lambda) / \bar{\varphi} - 1$, where $\bar{\varphi}$ is the mean transmitted flux fraction at the \textsc{Hi} absorber redshift, computed over  the entire sample.  We use here again a single \ion{H}{I} optical depth model, which is known to roughly match existing data:

\begin{equation}
\tau_{\rm eff} = \alpha \times (1+z)^{\beta} \;\;{\rm with}\,\alpha=0.0025\;{\rm and}\;\beta=3.7
\label{eqn:taueff}
\end{equation}

\noindent In practice, the Lyman-$\alpha$ flux power spectrum is inferred in the range adapted either to published SDSS spectra, $k=10^{-3} - 2\times 10^{-2}$~s km$^{-1}$, or to higher-resolution spectra, $k=10^{-3} - 0.1$~s km$^{-1}$. We exploit the splicing technique as described in~\citet{Borde2014}. It consists in combining the results of two N-body simulation outputs, one of high resolution with $768^3$ DM particles in a 25~$h^{-1}\,$Mpc box, and one on larger scales, with $768^3$ particles in a $100\,h^{-1}\,$Mpc box, making use of a third low-resolution simulation with $128^3$ particles in a 25~$h^{-1}\,$Mpc box.

Simulations were computed for four different values of $m_a$ between $3.4\times 10^{-22}$ and $4.1\times 10^{-21}$~eV using \texttt{AxionCAMB}. To assess FDM-related systematic effects, an additional simulation was run with $T(k)$ given by the formula from~\citet{Hu2000a}.

\section{Results}

Before comparing the predictions for the one-dimensional Lyman-$\alpha$ flux power spectrum with measured spectra, we first provide a discussion of the quantum pressure term, which was ignored in the N-body simulations. All the calculations presented here are therefore {\it a posteriori} and only hold if the dynamical impact of quantum pressure in the non-linear regime is negligible.

\begin{figure}
\includegraphics[width=\columnwidth]{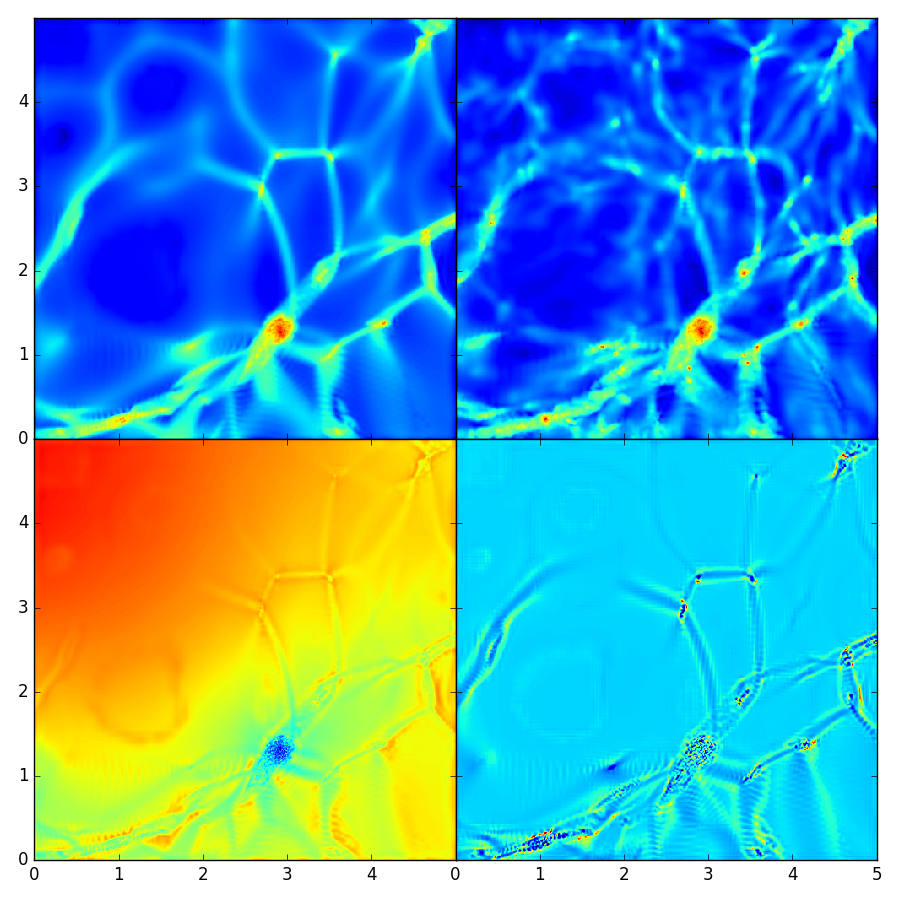}
\caption{Slice views of dark matter properties at $z=2.6$ from our ($768^3$ particles, $25 h^{-1}$~Mpc) simulations. Coordinates are in $h^{-1}$~Mpc. Top : comparison of the DM density fields for $m_a=3.4\times 10^{-22}$~eV (left) with respect to CDM (right). Bottom left : gravitational potential, with color scale in the range from $-4$ to $1\times 10^{10}$~(m/s)$^2$. Bottom right : quantum pressure for $m_a=3.4\times 10^{-22}$~eV, with color scale in the range from $-4$ to $8\times 10^{5}$~(m/s)$^2$.}
\label{fig:img_simus}
\end{figure}

\subsection{Quantum pressure}

Fig.~\ref{fig:img_simus} illustrates the properties of the DM fluid, derived from the \texttt{Gadget} snapshots for $z=2.6$ at scales of a few Mpc, which correspond to the median redshift and smallest comoving scales of relevance for the SDSS Lyman-$\alpha$ flux. The top panel provides a by-eye comparison of the DM mass density for CDM with respect to the lowest-mass FDM used in the N-body simulations, $m_a=3.4\times 10^{-22}$~eV. The severe attenuation of small-scale structures due to the linear cutoff in the FDM scenario is evident.

In order to assess the relative importance of quantum pressure, the bottom panel compares the gravitational potential $\phi$ (left), as calculated explicitly with \texttt{Gadget}, with the quantum pressure $Q/m_a$ (right). Both are expressed in $(m/s)^2$. The $Q$ term is estimated from the numerical laplacian of the density field $\rho$, which is itself obtained by smoothing the DM point particle distribution with a kernel adapted to the local density of DM particles in the simulation, so that higher resolutions are obtained in higher density regions. We checked that the resulting $Q$ distributions are stable with respect to the kernel size parameter, which means that our estimation for $Q$ is not severely biased by shot-noise related fluctuations. On the other hand, we find that the gradient estimator is limited by the simulation resolution in regions where the DM density fluctuates rapidly, such as inside halos of course, but also along the filament edges : this means that in these regions higher-resolution simulations would necessarily predict larger values for $Q$. With this caveat in mind, we can observe that $Q$ is significant only in small regions of space, again along filaments and in the vicinity of DM halos. The fact that gradients of $Q$ look concentrated along filaments is consistent with the results of~\citet{Veltmaat2016}, who find that the main differences in DM density maps when fully taking into account the $Q$ term are in filaments (appart from halo cores which they don't resolve). Note that, although it could be expected that $Q$ would blow up in deep voids due to the $1/\sqrt{\rho}$ factor~\citep{Marsh2015a}, we do not find any hint of this behavior from our calculations.

\begin{figure}
\includegraphics[width=\columnwidth]{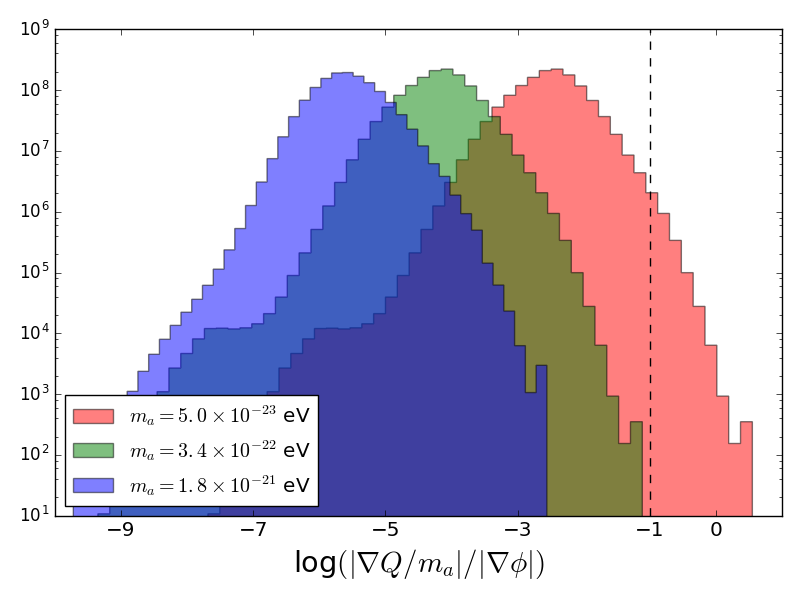}
\caption{Distributions of the ratio $r_Q$ between quantum and gravitational forces, for three different values of $m_a$. The distributions are weighted by DM mass and obtained from ($768^3$ particles, $25 h^{-1}$~Mpc) simulations at $z=2.6$. In the case $m_a=5.0\times 10^{-23}$~eV, the distribution is derived from a $m_a=3.4\times 10^{-22}$~eV snapshot, by simply rescaling the $Q$ term in $1/m_a$. The dashed line indicates the value $r_Q=10$~\%, above which quantum pressure becomes significant and should be incorporated in N-body simulations for consistency.}
\label{fig:forceratio}
\end{figure}

To quantify the validity of the N-body approximation used, we compute from \texttt{Gadget} snapshots the ratio between the two "force" terms which appear in the right-hand-side of Eqn.~\ref{eqn:euler}, based on our available estimate for $Q$:

\begin{equation}
r_Q \equiv  \frac{| \nabla Q / m_a |}{|\nabla \phi|}
\end{equation}

\noindent The mass-weighted distribution of this dimensionless ratio is represented on Fig.~\ref{fig:forceratio} for several values of $m_a$, using high-resolution snapshots at $z=2.6$.

For $m_a=1.8\times 10^{-21}$~eV (which is close to the limits that we will set in Section 3.3), and even for the lowest FDM mass implemented in our simulations, $m_a=3.4\times 10^{-22}$~eV, the quantum force \textit{at the scales resolved by our simulations} is always much smaller than the gravitational force. For $m_a=3.4\times 10^{-22}$~eV, we find that $r_Q>0.01$ for much less than 1~\% of the DM particles. The impact of $Q$ on the DM dynamics is therefore probably negligible. We observe a strong increase of $r_Q$ for higher redshifts due to the variation of gradients with the scale factor $a$, but that does not change the conclusion. However we remind again the caveat that our estimator for $Q$ is such that the quantum force is significantly underestimated in unresolved, high-gradient regions - and we know indeed from eg.~\citet{Schive2014}, that the quantum force does play a major role especially inside DM halos for such values of $m_a$. 

For values of $m_a$ significantly lower than $\sim 10^{-22}$~eV, such as $m_a=5.0\times 10^{-23}$~eV, Fig.~\ref{fig:forceratio} shows the distribution of $r_Q$ obtained by a simple $1/m_a^2$ rescaling with respect to the case $m_a=3.4\times 10^{-22}$~eV. We see that a significant population of FDM particles is expected to undergo a "quantum force" more than 10~\% times the gravitational force, mostly along filaments. This implies that the use of standard N-body simulations may be inappropriate in this mass regime, if one wants to predict density power spectra with high precision.

This discussion suggests that it is relevant to use classical N-body simulations in the context of this work, for scales resolved by our simulations (ie. larger than $\sim 100$~kpc), for FDM masses large enough, roughly $m_a \ga 10^{-22}$~eV. This is consistent with the results of~\citet{Veltmaat2016}, who find more explicitly that the effect of quantum pressure in the non-linear regime does not affect the power spectrum for modes $k\lesssim 100$~Mpc$^{-1}$ in the case $m_a=2.5\times 10^{-22}$~eV. Importantly, it is clear that the validity of this approximation is directly linked to the explored scales : the use of the relatively low-resolution SDSS Lyman-$\alpha$ flux power spectrum is therefore more robust against wave effects not accounted for, with respect to higher-resolution probes such as the XQ-100, HIRES and MIKE spectra.

\subsection{Lyman-$\alpha$ flux power spectra}

\begin{figure}
\includegraphics[width=\columnwidth]{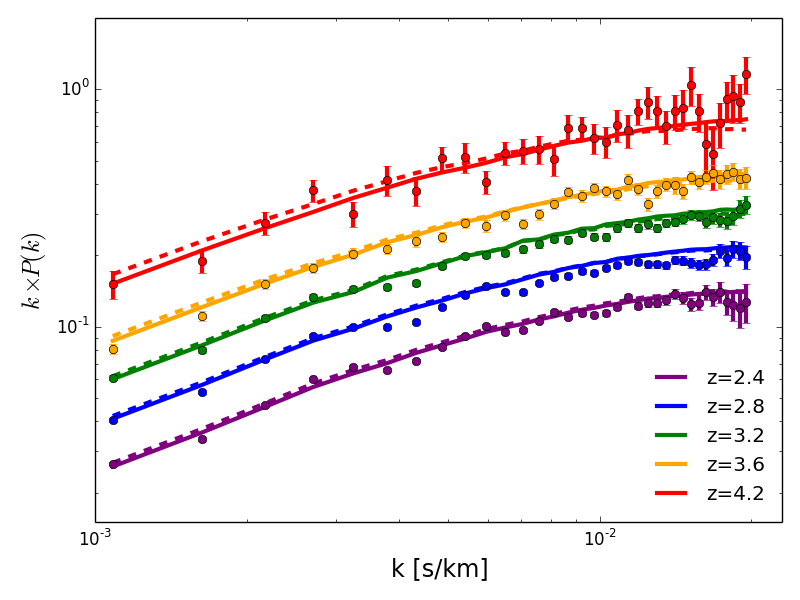}
\caption{Calculated one-dimensional Lyman-$\alpha$ flux power spectra from our benchmark CDM (solid line) and the $m_a=3.4\times 10^{-22}$~eV simulations (dashed). Overlaid for illustration are the SDSS-DR9 spectra at the corresponding redshifts. Note that the calculated fluxes do not fit the data, since the benchmark parameters used in the simulations are not the best fit results adjusted in Section 3.3, and no correction is applied to account for the \ion{Si}{iii}-\ion{H}{I} cross-correlation.}
\label{fig:pk1d_data}
\end{figure}

The Lyman-$\alpha$ flux power spectrum as calculated according to Section 2.3 at scales measured by SDSS is illustrated in Fig.~\ref{fig:pk1d_data}, for several redshifts of interest, both in the case of CDM and FDM with $m_a=3.4\times 10^{-22}$~eV. The difference between the predicted fluxes is rather subtle, of the order of 5~\%. This is however sufficiently large so that SDSS data can discriminate between models, using both the $k$ and $z$ signal dependences. On the contrary, for the highest FDM mass probed by our simulations, $m_a=4.1\times 10^{-21}$~eV, the typical differences in flux power with respect to CDM are below 1~\% and therefore hard to test with only SDSS spectra.

\begin{figure}
\includegraphics[width=\columnwidth]{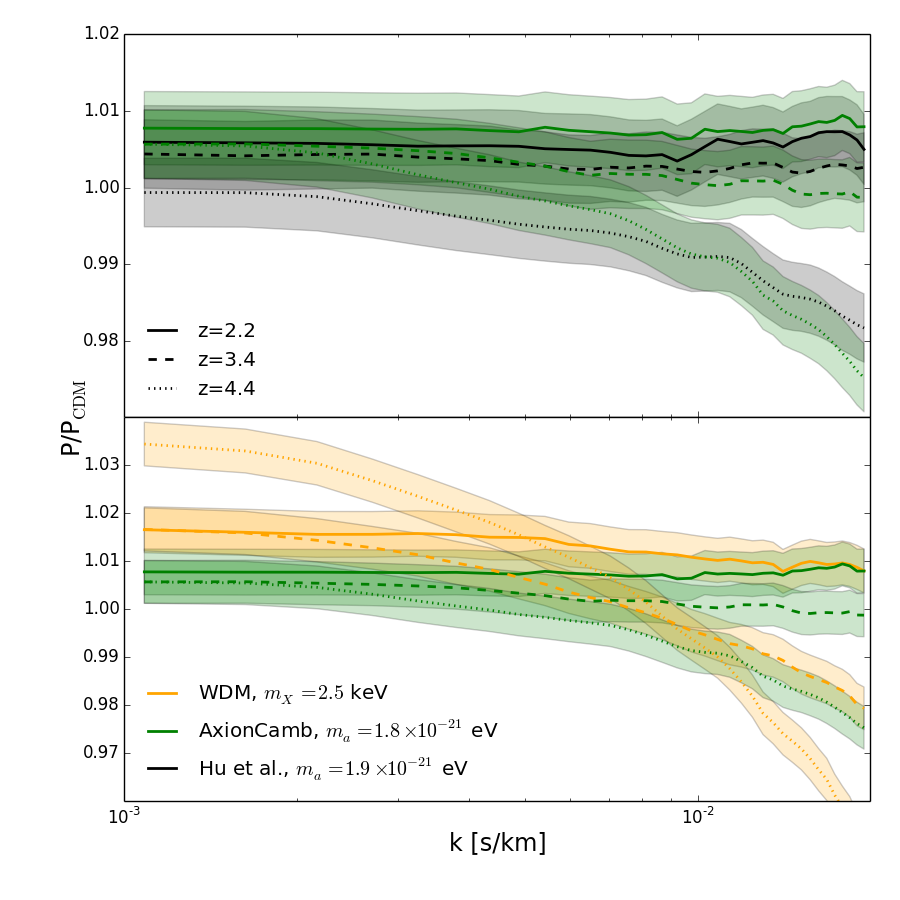}
\caption{Predicted ratio of one-dimensional Lyman-$\alpha$ flux spectra at different redshifts with respect to the equivalent CDM for different shapes of the linear $T(k)$. All models have the same linear cutoff $k_c$ such that $T(k_c)=0.5$. Top panel : comparison of the \texttt{AxionCAMB} prediction (green) with respect to the formula by~\citet{Hu2000a} (black). Bottom : \texttt{AxionCAMB} (green) versus WDM (orange). Note that these flux power spectra are normalized in order for all of them to obey the $\tau_{\rm eff}$ prescription given in Eqn.~\ref{eqn:taueff}.}
\label{fig:pk1d_cmpmodels}
\end{figure}

Fig.~\ref{fig:pk1d_cmpmodels} illustrates the impact of shape modifications in the linear transfer function $T(k)$ on the final Lyman-$\alpha$ flux prediction, for transfer functions with the same cutoff value $k_c$. While the differences between the linear power spectra predicted by \texttt{AxionCAMB} and the Hu formula are not negligible at all, as discussed in Section 2.1, the top panel of Fig.~\ref{fig:pk1d_cmpmodels} shows that the related difference for the SDSS Lyman-$\alpha$ spectra is well below 1~\%. Therefore we conclude that the uncertainties related to the modeling of the scalar field linear evolution are negligible for Lyman-$\alpha$ forest predictions.

On the contrary, the shape difference between FDM and WDM transfer functions do propagate significantly to the Lyman-$\alpha$ forest, as Fig.~\ref{fig:pk1d_cmpmodels} (bottom) demonstrates. This is due to the fact that the WDM transfer function is smaller than the FDM one for values of $k$ much smaller than the cutoff $k_c$. Finally, let us observe from Fig.~\ref{fig:pk1d_cmpmodels} that the flux power spectrum differences are more pronounced at high redshift, while they are reduced at low $z$ where non-linear effects tend to erase the original shape differences.

The Lyman-$\alpha$ flux power spectrum is also attenuated by the thermal pressure which smooths the spatial distribution of IGM over cosmological times. The relevant Jeans smoothing scale is given by~\citep{Schaye2001} $L_J \simeq 0.52\,{\rm kpc}\times (T/10^4{\rm K})^{1/2} \, (n_H/{\rm cm}^{-3})^{-1/2}$. Using the values for $T$ and $n_H$ calculated in our CDM simulation, the spatial distribution of $L_J$ is peaked around $L_J\sim 500$~kpc. The associated mode $2\pi/L_J$ corresponds to the value $m_a \simeq 7\times 10^{-22}$~eV, using Eqn.~\ref{eqn:fdm-cutoff}. The effect of thermal Jeans smoothing happens therefore at a scale comparable to the one associated to the FDM effect we're looking for. There is therefore a degeneracy between IGM temperature parameters $(T_0,\gamma)$ and $m_a$.

\subsection{Constraining $m_a$ from the SDSS/BOSS Lyman-$\alpha$ forest}

The SDSS Lyman-$\alpha$ forest data used here is the same as in~\citet{Palanque-Delabrouille2013a}, in which all the computation details for the flux power spectrum and its related statistical and systematic uncertainties are extensively described. From a parent sample of $\sim 60,000$ SDSS-III/BOSS DR9 quasars~\citep{Ahn2012a, Eisenstein2011, Ross2012}, the data consist in a selection of $13,821$ spectra that have signal-to-noise ratios per pixel greater than 2, no broad absorption line features, no damped or detectable Lyman-limit systems, and an average resolution in the Lyman-$\alpha$ forest of at most  $85 \: \rm km~s^{-1}$. The Lyman-$\alpha$ forest is defined as the region spanning  $1050 < \lambda_{RF} / \r{A} < 1180$. The spectra in this sample are used to  measure the transmitted flux power spectrum in 12~redshift bins from $\langle z \rangle = 4.4$ to $2.2$, each bin spanning $\Delta z = 0.2$, and in 35 equally-spaced spatial modes ranging from $k=10^{-3}$ to $2\times 10^{-2} ~\rm s~km^{-1}$. The flux power spectrum is obtained from the Fourier Transform of the fractional flux transmission $\delta_\varphi$ defined in Section 2.3, computed separately for each $z$-sector. 

It has been shown (eg. in~\citet{Palanque-Delabrouille2013a} that this SDSS Lyman-$\alpha$ forest matches well predictions based on the CDM model, by appropriately choosing relevant astrophysical and instrumental parameters such as the IGM thermal state modeling or the spectrometer resolution. We can therefore constrain the $m_a$ parameter of FDM in a way similar to constraints previously set on the sum of active neutrino masses, or the mass $m_X$ of WDM, using the same data. We use the same likelihood as used in~\citet{Baur2015, Palanque-Delabrouille2015a}, in order to handle the influence of relevant cosmological and astrophysical parameters : $H_0$, $\Omega_M$, $n_s$ and $\sigma_8$ on the one hand, and on the other hand the previously mentioned parameters $(T_0,\gamma)$ (Eqn.~\ref{eqn:t_igm}) and $(\alpha,\beta)$ (Eqn.~\ref{eqn:taueff}), which provide a simple description of the IGM temperature and mean Lyman-$\alpha$ optical depth. No CMB-based prior is set for the cosmological parameters, except for $H_0$, to which we apply the Gaussian constraint $H_0=67.3\pm1$. No (ie. flat) priors are used for $(\alpha,\beta)$, while we constrain $\gamma=1\pm0.3$ and $T_0\in [0;25000]$~K.

Other physical processes may impact significantly our FDM constraints, but are poorly known and not explicitly modeled in our simulations. To capture their effects, we introduce additional nuisance parameters, with simple analytical modifications of the predicted flux power spectra, based on other published simulations or models. The description of these corrections is detailed in~\citet{Baur2015}, and they include feedback processes from galactic outflows~\citep{Viel2013a}, fluctuations of the UV background from discrete sources, and homogeneous reionization history~\citep{McDonald2005}.
Other identified nuisance parameters are taken into account, to model the \ion{Si}{iii}-\ion{H}{I} cross-correlation, the spectrometer noise, the presence of residual damped Lyman-$\alpha$ systems in the data, and simulation uncertainties related to the splicing algorithm.
The signal dependence with $m_a$ is captured using the grid of simulations with $m_a=3.4\times 10^{-22}, 7.9\times 10^{-22}, 1.8\times 10^{-21}$ and $4.1\times 10^{-21}$~eV, and for which all other parameters are fixed to the benchmark values given in Section 2. For each FDM simulation a $\chi^2$ is computed with respect to the $35\times 12$ SDSS data points in $(k,z)$ space, assuming $h=0.673\pm0.010$~\citep{PlanckCollaboration2015} while floating all other likelihood parameters.

We then explored two methods to derive a bound on $m_a$ from these $\chi^2$s. In a first approach, we use already existing CDM simulations which vary all the aforementioned astrophysical and cosmological parameters while keeping $1/m_a=0$ in order to extrapolate the grid at any point of the whole parameter space. Since no FDM simulation including cross terms between $m_a$ and other parameters was carried out, this is done by assuming that the variation of the flux power spectrum with all parameters at any given value of $m_a$ is identical to the one calculated for $1/m_a=0$.  We then identify the minimal $\chi^2$ value, letting all parameters free. It occurs for $\sigma_8=0.85$, $n_s=0.94$, $\Omega_M=0.29$,  $(T_0,\gamma)=(8723,0.93)$, $(\alpha,\beta)=(0.0025,3.7)$, and $1/m_a=0$ - meaning that the best fit is the pure CDM scenario. The minimal $\chi^2$ is 408.1, comparable to the one for the highest tabulated FDM mass, $\chi^2(m_a=4.1\times 10^{21}~{\rm eV})=409.0$. For the tabulated FDM model with lowest mass we find $\chi^2(m_a=3.4\times 10^{-22}~{\rm eV})=431.6$, which shows that such a value for $m_a$ is clearly excluded. A frequentist 95~\% lower limit on $m_a$ is set, following eg.~\citet{Palanque-Delabrouille2014} : $m_a> 2.0\times 10^{-21}$~eV. As already mentioned in~\citet{Baur2015}, all adjusted best-fit IGM nuisance parameters are compatible with no correction within $1\sigma$.

In order to by-pass the lack of FDM cross term calculations, and also to highlight the correspondance with WDM models, we also used a second approach which consists in using the WDM bound published in~\citet{Baur2015} in the following way. We determine again a relationship between $m_X$ and $m_a$, this time so that the same $\chi^2$ is obtained in the SDSS adjustment, $\chi^2(m_X) = \chi^2(m_a)$. This requirement is satisfied with the following scaling law:

\begin{equation}
m_X = 0.715\times \left(\frac{m_a}{10^{-22}\,{\rm eV}}\right)^{0.558}\;{\rm keV}
\label{eqn:mX-ma-fullscaling}
\end{equation}

\noindent This scaling is significantly different from Eqn.~\ref{eqn:mass-scaling-2} which was derived from the cutoff position in linear matter spectra. However, it is the best adapted for our goal in the sense that the WDM and FDM models associated by this relation do predict Lyman-$\alpha$ flux power spectra that are indiscernible using current SDSS data. The 95~\% CL limit $m_X>4.09$~keV derived by~\citet{Baur2015} then converts to $m_a>2.3\times 10^{-21}$~eV. Given that this second approach takes into account cross terms, calculated in~\citet{Baur2015}, we consider it to be more accurate than the first one, and therefore conclude that the SDSS Lyman-$\alpha$ forest exclude FDM models with $m_a<2.3\times 10^{-21}$~eV at 95~\% CL.

The above-mentioned matter spectral index adjusted with this SDSS data only, $n_s=0.94$, is in slight tension with the CMB-derived value $n_s=0.97$~\citep{PlanckCollaboration2015}. As was noticed in~\citet{Baur2015}, using SDSS only without Planck-based priors (except for $H_0$) is similar to allowing for a running of $n_s$ between the large scales of CMB and the smaller Lyman-$\alpha$ regime. Applying Planck-based priors on all our cosmological parameters therefore leads to a less tight constraint on the FDM or WDM mass. The 28~\% change in $m_X$ derived in~\citet{Baur2015} translates into the looser bound $m_a>1.3\times 10^{-21}$~eV if one includes~\citet{PlanckCollaboration2015} into the adjustment.

Uncertainties linked to the modeling of FDM itself were already extensively discussed. We noticed that uncertainties in the linear matter spectrum do not modify the result. The impact of quantum pressure in the N-body simulations is more an open question, but we argued that at least for $m_a \ga 10^{-22}$~eV it should be negligible at the scale and precision probed by the SDSS Lyman-$\alpha$ forest.

Finally, errors related to the simplified treatment of IGM physics in our model were taken into account by allowing to float "nuisance" parameters, from $(T_0,\gamma)$ to the analytical corrections for feedback effects or UV fluctuations. However, these corrections are derived from simulations or models with specific underlying assumptions, so that it cannot be certain that they fully take into account all processes relevant for our study. In addition, other phenomena that may impact the \ion{H}{I} gas properties were not taken into account, such as residual fluctuations due to the inhomogeneity of the reionization process~\citep{Cen2009}. This specific process should however affect more the high-redshift IGM properties.

\subsection{Adding higher-resolution Lyman-$\alpha$ forest data}

The analysis presented in Section 3.3 can easily be extended by including additional Lyman-$\alpha$ forest data with higher resolution. For that purpose, in this Section we add to the SDSS flux power spectrum the data from both the XQ-100 survey~\citep{Lopez2016} and HIRES/MIKE data as presented in~\citet{Viel2013}.

The XQ-100 survey, obtained with the XSHOOTER spectrograph at VLT, consists of a homogeneous and high-quality sample of 100 quasar spectra at $z=3.5 - 4.5$. The Lyman-$\alpha$ flux power spectrum was first derived from this survey in~\citet{Irsic2016a}, and the XQ-100 publicly available raw data was recently reanalyzed by~\citet{Yeche2017a}, taking into account in particular an improved determination of the spectrograph resolution. Here we use the Lyman-$\alpha$ flux power spectrum as tabulated in~\citet{Yeche2017a}, for three values of average redshifts in the range $z=3.2 - 3.93$, and spatial modes reaching values as high as $k=0.05 - 0.07$~s km$^{-1}$, depending on the redshift. 

In addition to XQ-100, we also include the Lyman-$\alpha$ flux power spectrum as presented in~\citet{Viel2013} (Fig. 7), which is based on a collection of quasar spectra from Keck/HIRES and Magellan/MIKE. These high-resolution spectrometers permit to reach spatial modes up to $k=0.09$~s km$^{-1}$. Here we use only the flux power spectra determined at $z \leq 4.6$ due to the lack of simulation snapshots available at higher redshifts.

With the same methodology as in Section 3.3, we combine the SDSS Lyman-$\alpha$ spectra with those higher-resolution data, as was done in~\citet{Yeche2017a}. We make use of the same simulated FDM flux power spectra, described in Section 2. With the procedure
based on the $m_X - m_a$ mass matching, the limit $m_X > 4.65$~keV from~\citet{Yeche2017a} translates into the following bound $m_a>2.9 \times 10^{-21}$~eV at 95~\% CL. This corresponds to a 30~\% improvement with respect to the limit based on SDSS only : the higher resolutions of XSHOOTER, HIRES and MIKE permit to explore fluctuations at smaller comoving lengths.

If we include Planck priors on the cosmological parameters as was done with SDSS alone, the 95~\% CL bound becomes $m_a>2.4\times 10^{-21}$~eV. The relative change in the constraint is less severe than with SDSS data alone. This is because the inclusion of higher-resolution spectra to the original SDSS data also reduces the above-mentioned tension on the $n_s$ parameter, as can be see in~\citet{Yeche2017a}.

The uncertainties which may impact this bound are similar to those discussed in the end of Section 2.3. The impact of potential small-scale fluctuations in the IGM properties related to the distribution of ionizing sources and patchy reionization is increased, although our measurements do not rely on observations at $z\geq 5$. Also, for a given value of $m_a$, the effect of quantum pressure is stronger at the scales probed by high-resolution spectrographs : the "quantum force", being a third-order derivative, scales as $\sim k^3$. However, while this effect could be problematic for $m_a\sim 10^{-22}$~eV, it should not be a concern for the mass range $m_a=2.3 - 2.9\times 10^{-21}$~eV which is not covered by SDSS alone.

\section{Conclusions}

In this article, we studied the Lyman-$\alpha$ forest phenomenology within the Fuzzy Dark Matter scenario. The non-linear physics and hydrodynamics were modeled using standard N-body simulations with initial conditions which reproduce the FDM linear physics. This approach is an approximation which neglects the "quantum force" present in the Madelung equation. We presented semi-quantitative arguments, based on the simulation outputs, which strongly suggest that at least for $m_a\ga 10^{-22}$~eV the quantum force at the scales resolved by our simulations is always much smaller than the gravitational force. In agreement with the conclusions of~\citet{Veltmaat2016}, we find indications that for $m_a\la10^{-22}$~eV the quantum term may impact the non-linear predictions for $P(k)$, and should therefore be included in the simulations in order to provide more reliable predictions.

Our main result is obtained by comparing the SDSS Lyman-$\alpha$ flux spectra and simulations, which permits to derived the limit $m_a>2.3\times 10^{-21}$~eV (95~\% CL). Including higher-resolution spectra from the XSHOOTER, HIRES and MIKE spectrographs extends the exclusion range up to $m_a=2.9\times 10^{-21}$~eV. This analysis was done with a similar method, and similar assumptions, as for the case of WDM models~\citep{Baur2015, Yeche2017a}. For this study we also derived the most appropriate $m_a - m_X$ scaling law adapted to the SDSS Lyman-$\alpha$ data, which is different from the one which was previously found using the linear transfer function cutoffs.

These constraints confirm that the Lyman-$\alpha$ forest provides sensitivity to FDM masses as large as or larger than other observables such as high-redshift galaxy luminosity functions~\citep{Menci2017}. More importantly, the Lyman-$\alpha$ forest is an independent probe with different sources of observational and astrophysical uncertainties. Together with others, our result now provides severe constraints on models tailored to solve small-scale problems of the CDM paradigm, for which $m_a \la 10^{-21}$~eV seems to be necessary.

While in the process of final redaction of this article, we were informed of the work presented in~\citet{Irsic2017a}. In their study, the authors of this work focus on the high-resolution Lyman-$\alpha$ forest to derive a bound on $m_a$. Their conclusions are very similar to ours in terms of constraints for $m_a$. In particular, our $m_X - m_a$ scaling law given by Eqn.~\ref{eqn:mX-ma-fullscaling} lies in between the curves quoted "$k_{1/2}$" and "$k_{0.75}$" in their Fig.~3, and interpolates reasonably their result points given by an MCMC analysis. We point out the complementarity in terms of systematics between both works. Whereas the work of~\citet{Irsic2017a} uses exclusively high-resolution and high signal-to-noise Lyman-$\alpha$ spectra, including at redshifts above $z=5$, their statistical uncertainties are larger, and we note that some astrophysical uncertainties associated with the physics of the IGM and the effect of reionization are more important at  high redshift. We also stress that the results of~\citet{Irsic2017a}, like ours, could be impacted by wave effects in the non-linear regime for $m_a\la 10^{-22}$~eV.

In addition to progress in the modeling of IGM physics, which is needed to improve the robustness of all Lyman-$\alpha$-based studies, future work should also be devoted to study the impact of non-linear wave effects for $m_a \la 10^{-22}$~eV. Baring this issue, our results provide among the best absolute lower bounds for the mass of bosonic dark matter. Even given this constraint, a number of interesting phenomena are predicted in the FDM framework that are distinctive from CDM and deserve to be explored in the future. More elaborate, but plausible scenarios such as mixed CDM-FDM models or models with mass-varying FDM are also worth investigating.

\section*{Acknowledgements}

We acknowledge fruitful discussions with Jim Rich on the physics of FDM, and with Christophe Magneville on the N-body simulation pipeline. We thank Daniel Grin for exchanges on the \texttt{AxionCAMB} program. The authors are also grateful to Vid Irsic and Matteo Viel for providing the HIRES/MIKE-derived flux power spectra.

We thank Volker Springel for providing us with the \texttt{Gadget-3} program. \texttt{Gadget} snapshots were exploited using the \texttt{pynbody} software~\citep{pynbody}.
We acknowledge PRACE (Partnership for Advanced Computing in Europe) for access to thin and xlarge nodes on the Curie cluster based in France at the TGCC (Tr\`es Grand Centre de Calcul) under allocation numbers 2010PA2777, 2014102371 and 2012071264. We also acknowledge the French national access to high-performance computing GENCI (Grand Equipement National de Calcul Intensif) for access to the Curie cluster under allocation t2016047706.
DJEM acknowledges support of a Royal Astronomical Society Postdoctoral Fellowship, hosted at King's College London.

\bibliographystyle{mnras}
\bibliography{FDM_Lyalpha} % if your bibtex file is called example.bib

\bsp	% typesetting comment
\label{lastpage}
\end{document}